\documentclass[twoside,fleqn]{article}
\usepackage{espcrc2}
\usepackage{epsfig}
\usepackage{graphicx}

\def\lsim{\raise0.3ex\hbox{$<$\kern-0.75em\raise-1.1ex\hbox{$\sim$}}}
\def\gsim{\raise0.3ex\hbox{$>$\kern-0.75em\raise-1.1ex\hbox{$\sim$}}}

\title{The Weak-Coupling Limit of $3D$ Simplicial Quantum Gravity}

\author{
  P.~Bialas$^{\rm a,b}$, B.~Petersson$^{\rm a}$ and 
  G.~Thorleifsson\address{
		   Fakult\"{a}t f\"{u}r Physik, Universit\"{a}t 
                   Bielefeld, 33501 Bielefeld, Germany \\
		   $^{\rm b}\;$Institute of Comp.\ Science, Jagellonian
                   University, 30072 Krakow, Poland}
}		             

\begin{document}

\begin{abstract}
We investigate the weak-coupling limit, $\kappa \rightarrow \infty$,
of $3D$ simplicial gravity using Monte Carlo simulations and
a Strong Coupling Expansion.  With a suitable modification
of the measure we observe a transition from a branched
polymer to a crinkled phase.  However, the
intrinsic geometry of the latter appears similar to that of non-generic
branched polymer, probable excluding the existence of a sensible 
continuum limit in this phase.
\end{abstract}

\maketitle

\section{INTRODUCTION}

$D$--dimensional simplicial quantum gravity is a discretization
of Euclidean quantum gravity with the integration over
space-time metrics replaced by a sum over all possible
$D$--dimensional triangulations constructed by gluing together
equilateral simplexes.  It is defined by the partition function
\begin{equation}
  Z(\mu,\kappa) \;=\;
     \sum_{N_D} \; e^{-{\mu} N_D} \;
    Z({\kappa},N_D) 
\end{equation}
\vspace{-12pt}
\[ \nonumber
 \hspace{37pt}
 = \sum_{N_D,N_0} \; e^{-{\mu} N_D + {\kappa} N_0}
 \;W_D(N_0,N_D) \;,\nonumber
\]
\vspace{-13pt}
\begin{equation} 
 W_D(N_0,N_D) \;= \sum_{T\in{\cal T}(N_0,N_D)} \frac{1}{C_T}\;,
\end{equation}
\vspace{-5pt}
where $N_D$ = \# $D$--simplexes, $N_0$ = \# vertices,
and $C_T$ is the {\it symmetry} factor of a labeled
triangulation $T$ chosen from a suitable
ensemble $\cal T$ ({\it eg} combinatorial).
$\mu$ and $\kappa$ are the discrete cosmological
and Newton's coupling constants.

In $D = 3$ and 4 this model has {\it two} phases:
\begin{itemize}
 \vspace{-5pt}
 \item{$\kappa < \kappa_c$} :
  a (intrinsically) {\it crumpled} phase
 \vspace{-5pt} 
 \item{$\kappa > \kappa_c$} :
  a {\it branched polymer} phase
\end{itemize}
\vspace{-5pt}
separated (regrettably) by a {\it discontinuous} phase transition
\cite{gen}.

As a discontinuous phase transition excludes a sensible
continuum limit, there have been
several attempts to modify the model Eq.~(1) in the hope of 
finding a non-trivial phase structure.
This includes: adding a {\it measure} term \cite{enso}
\vspace{-5pt}
\begin{equation}
 W_D(N_0,N_D,{\beta}) \;= \sum_{T\in
 {\cal T}(N_0,N_D)} \frac{1}{C_T} \;
    {\prod_{i=0}^{N_0} q_{i}^{{\;\beta}}},
\end{equation}
\vspace{-2pt}
where $q_i$ is the order of the vertex $i$ (number of simplexes
containing $i$),
and coupling matter fields to the geometry \cite{bilke}.

Such modifications do indeed lead to a more complicated 
phase diagram (Fig.~1), and suitable modified the 
model exhibits a new {\it crinkled} phase.
But does this new phase 
structure imply a more interesting non-trivial
critical behavior?
To investigate this we have studied the weak-coupling
limit, $\kappa \rightarrow \infty$, of the model Eq.~(1)
for $D=3$.
\begin{figure}[t]
 \centerline{\includegraphics[width=2.7in,bb=57 270 573 590]{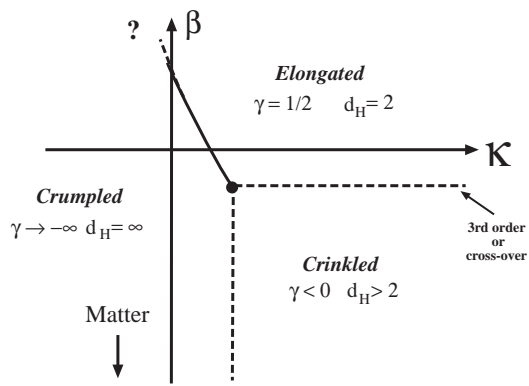}}
 \label{fig1}
 \caption{A schematic phase diagram of simplicial gravity
  in 3 and 4 dimensions.}
\end{figure}

\section{THE EXTREMAL ENSEMBLE}    

In the weak-coupling limit the partition function Eq.~(1),
in $D=3$ and 4, is expected to be dominated by an 
Extremal Ensemble (EE) of triangulations.
For this ensemble, defined as triangulations with
the maximal ratio $N_0/N_D$, the partition function simplifies:
\begin{equation}
 Z(\mu) \;\;=\; \sum_{N_D} \; {\rm e}^{- \mu N_D}
     \;W_D(N_D)
\end{equation}
\vspace{-4pt}
\[   
  W_D(N_D) \;= \hspace{-10pt} \sum_{T\in
  {\cal T}(N_0^{\rm max},N_D)} \;\frac{1}{C_T}
\]
where \vspace{-2pt} 
\begin{equation}
    N_0^{\rm max} \;=\;  \left \{ \begin{array}{cc}
     \left\lfloor  \frac{\textstyle N_3+10}{\textstyle 3}\right\rfloor
      & D = 3\;,  \\ \vspace{-5pt} \\
    \left\lfloor \frac{\textstyle N_4+18}{\textstyle 4}\right\rfloor
      & D = 4\;. 
    \end{array} \right.
\end{equation}
Here $\lfloor x \rfloor$ denotes the floor function ---
the biggest integer not greater than $x$.
This in turn defines several {\it distinct} series
for the EE:
\[
  3D \; : \;\;
   S^0\; \bigl (N_0,\;3 N_0\!-\!10\bigr ),
    \;\;  S^1\;\bigl (N_0,\;3 N_0\!-\!9\bigr ), 
\]
\vspace{-14pt}
\[
   \hspace{30pt} S^2\;\bigl (N_0,\;3 N_0\!-\!8\bigr ).
\]
\vspace{-12pt}
\[ 4D \; : \;\;
   S^0\;\bigl (N_0,\;4N_0\!-\!18\bigr ),
    \;\;\; S^1\;\bigl (N_0,\;4 N_0\!-\!17\bigr ).
\]

Assuming the asymptotic behavior $W_D(N_D) \sim
\exp (-\mu_c N_D)\;N_D^{\gamma - 3}$, which defines the
string susceptibility exponent $\gamma$,
we observe (from a SCE) that for the different series $S^k$,
$\gamma^k=k+\frac{1}{2}$.  This difference in the 
exponent $\gamma$ can be understood as the 'higher' series
($k = 1,2,\ldots$) can be constructed by introducing
$k$ ``defects'' (marked points) into triangulations
belonging to the minimal series $S^0$.
Moreover, we observe that
the minimal series appears to have very small finite-size effects.

The minimal series $S^0$ can be explicitly {\it enumerated}
as it corresponds to $D$--dimensional 
{\it combinatorial stacked spheres}
(CSS), {\it ie} to the  surface of
a $(D+1)$--dimensional simplicial cluster.
The number of $(D+1)$--dimensional simplicial clusters build
out of $n \; (D+1)$ simplexes, rooted at a marked outer face,
is given by (where $n=N_0-D-1$) \cite{her}
\[
 e_{D+1,n} \;=\!\!\!\!\!\!
     \sum_{\scriptscriptstyle \begin{array}{c}
             n_1+\cdots +n_{D+1}\\
             =n-1
             \end{array}}
      \!\!\!\!\!e_{D+1,n_1} \cdots e_{D+1,n_{D+1}} 
\]
\vspace{-6pt}
\begin{equation}
      \hspace{35pt}=\; \frac{1}{nD +1 }\;\left (
          \begin{array}{c}
           (D\!+\!1)\; n \\
           n
           \end{array} \right )
\end{equation}

\begin{equation}
 \Rightarrow  \;\;\  W_{D}(N_D) \;= \; \frac{D+2}{N_D} \;
     e_{D+1,\frac{N_D-2}{D}}\;.
\end{equation}
Expanding this gives
\vspace{-5pt}
{\small
\[
 W_{3}(N_3) = \frac{10}{\sqrt{2\pi} \; N_3^{5/2}}
   \left (\frac{256}{27}\right)^{\frac{N_3-2}{3}}
   \left ( 1
      +\frac{83}{48}\frac{1}{N_3} \right.
     \cdots   
\]
\vspace{-10pt}
\[
   W_{4}(N_4) = \frac{6\sqrt{5}}{\sqrt{2\pi}\;N_4^{5/2}}
   \left (\frac{3125}{256}\right)^{\frac{N_4-2}{4}}
   \left ( 1+\frac{33}{20}\frac{1}{N_4}  \right.
     \cdots  
\]
} 
\noindent
with $\gamma = 1/2$ as expected for branched polymers.

\section{A MODIFIED MEASURE}

\begin{figure}[t]
 \centerline{\includegraphics[width=2.7in,bb=104 250 465 589]{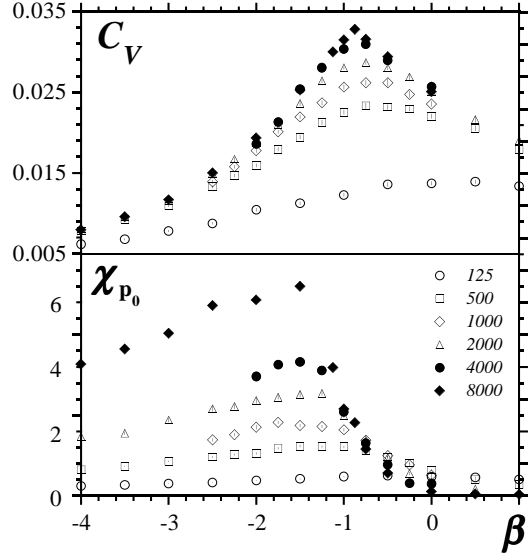}}
 \label{fig2}
 \caption{Evidence of a phase transition in the
 $3D$ EE, Eq.~(4), with a modified measure Eq.~(3)
 ({\sc Top}) The fluctuations in the measure term: $C_V$, 
  and ({\sc Bottom}) in the maximal vertex order: $\chi_{p_0}$.}
\end{figure}

We have investigated the $3D$ EE
including a measure term, using both
MC simulations and a SCE \cite{us}.
We find a {\it continuous} phase transition to a
crinkled phase at $\beta \approx -1$ (Fig.~2).
This is evident in the fluctuations both in
the measure term --- the ``specific heat'' $C_V$ ---
and in the maximal vertex order $p_0$.  
Scaling analysis of the peak value of 
specific heat gives: $C_V^{\rm max} \approx
a + b N_3^{-0.34(4)}$.
 
To explore the fractal properties of the geometry
in the crinkled phase  
we have measured the variations in $\gamma$ 
with $\beta$ using several different methods (Fig.~3).  
As in $D = 4$, we find that $\gamma$
becomes negative at $\beta_c$ and decreases 
with $\beta$.  Similarly we find a spectral
dimension that increase from $d_s = 4/3$ 
for $\beta > \beta_c$, to $d_s \approx 2$ as 
$\beta \rightarrow \infty$.
 
Estimates of the intrinsic
fractal dimension $d_H$ differ, on the other hand, 
substantially depending on
how it is defined --- on the direct graph
(from a vertex-vertex distribution) or on 
the dual graph (simplex-simplex distribution).
The former yields $d_H \rightarrow \infty$,
the latter $d_H \approx 2$. 
In addition, we observe that the crinkled phase
appears dominated by a {\it gas} of sub-singular vertices.

Combined this evidence suggests that the crinkled phase probably
corresponds to some kind of non-generic branched polymers phase which
makes it unlikely that any sensible continuum limit exist in this
phase.  This of course does not exclude the possibility that a second
phase order transition point exists somewhere on the phase boudary,
for example at the end of the first order transition line (Fig.~\ref{fig1}). 
\begin{figure}[t]
 \centerline{\includegraphics[width=2.7in,bb=122 300 479 547]{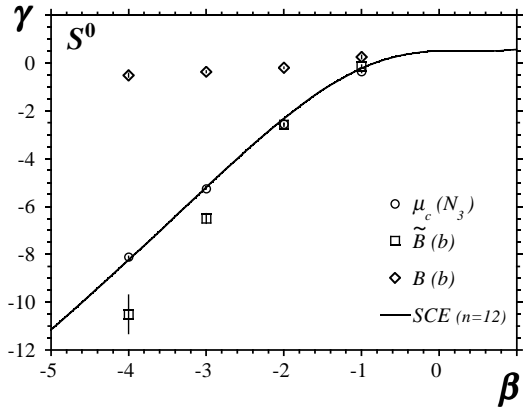}}
 \label{fig3}
 \caption{Variations in $\gamma$ with $\beta$
  for the EE Eq.~(4) with a modified measure
  Eq.~(3).}
\end{figure}

\section{DEGENERATE TRIANGULATIONS}

The EE can also be defined with the ensemble
of {\it degenerate} triangulations 
introduced in Ref.\ \cite{deg}.
In this case degenerate stacked spheres (DSS)
are constructed by slicing open a face and inserting 
a vertex.  Different from CSS, Eq.~(5), DSS 
are defined by the maximal ratio:
\begin{equation} 
 \frac{N_0}{N_D} \;=\; \frac{1}{2}+\frac{D}{N_D}.
\end{equation}
This ensemble can also be enumerated explicitly.

Modifying the measure leads to identical phase
structure as is observed for CSS.  This
is shown in Fig.~4 where we plot the 
variations in $\gamma$ with $\beta$ for the 
two ensembles.  That the two, very different,
ensembles agree on the fractal 
structure is reassuring and reflects the universal
properties of the crinkled phase.
\begin{figure}[t]
 \centerline{\includegraphics[width=2.7in,bb=122 300 479 547]{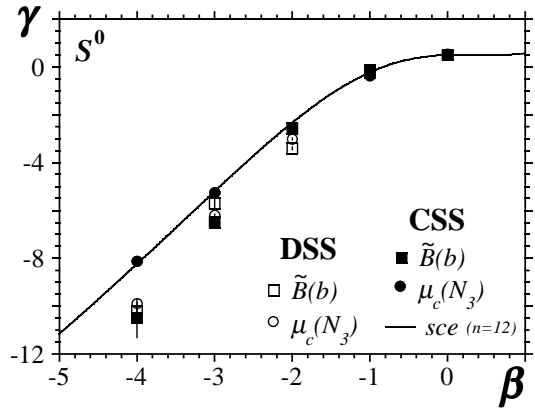}}
 \caption{\label{fig4}bel{fig4}Variations in $\gamma$ with a modified
  measure, both for an ensemble of DSS and 
  CSS.}
\end{figure}

{\bf Acknowledgments} P.~B. was supported by the Alexander von Humboldt
Foundation.


\begin{thebibliography}{99}
 
\bibitem{gen} 
 G.~Thorleifsson, {\it Nucl.~Phys}.\ {\bf 73} (Proc.\ Suppl).\
     {\bf 73} (1999) 133.
     
\bibitem{enso}
 B.~Brugmann and E.~Marinari, {\it Phys.\ Rev.\ Lett}.\
  {\bf 70} (1993) 1908.

\bibitem{bilke}
 S.~Bilke, {\it et al}, {\it Phys.~Lett}.\ {\bf B418}
  (1998) 266; {\bf B432} (1998) 279.
     
\bibitem{her}
 F.~Hering, R.C.~Read and G.C.~Shephard, 
 {\it Discr.\ Math}.\ {\bf 40} (1982) 203.

\bibitem{us}
 G.~Thorleifsson, P.~Bialas and B.~Petersson, {\it Nucl.~Phys}.\
     {\bf B550} (1999) 465.
     
\bibitem{deg}
 G.~Thorleifsson, {\it Nucl.~Phys}.\ 
  {\bf B538} (1999) 278.
  
\end{thebibliography}
\end{document}